\documentclass[preprint,journal]{vgtc}            

\preprinttext{To appear in IEEE Transactions on Visualization and Computer Graphics.}

\renewcommand{\preprintmanuscriptnotetxt}{\vspace{30pt}} 

\vgtccategory{Research}

\vgtcpapertype{Representations \& Interaction}

\title{A Computational Design Pipeline \\to Fabricate Sensing Network Physicalizations}

\author{%
 S. Sandra Bae, Takanori Fujiwara, Anders Ynnerman,\\ Ellen Yi-Luen Do, Michael L. Rivera, and Danielle Albers Szafir
}

\authorfooter{
  \item
  	S. Sandra Bae, Ellen Yi-Luen Do, Michael L. Rivera are with University of Colorado, Boulder.
  	E-mail: \{sandra.bae, ellen.do, mrivera\}@colorado.edu
  \item
  	Takanori Fujiwara, Anders Ynnerman are with Link\"{o}ping University.
  	E-mail: \{takanori.fujiwara, anders.ynnerman\}@liu.se

  \item Danielle Albers Szafir is with University of North Carolina-Chapel Hill.
  	E-mail: danielle.szafir@cs.unc.edu.
}

\abstract{Interaction is critical for data analysis and sensemaking. However, designing interactive physicalizations is challenging as it requires cross-disciplinary knowledge in visualization, fabrication, and electronics. Interactive physicalizations are typically produced in an unstructured manner, resulting in unique solutions for a specific dataset, problem, or interaction that cannot be easily extended or adapted to new scenarios or future physicalizations. To mitigate these challenges, we introduce a computational design pipeline to 3D print network physicalizations with integrated sensing capabilities.
Networks are ubiquitous, yet their complex geometry also requires significant engineering considerations to provide intuitive, effective interactions for exploration. Using our pipeline, designers can readily produce network physicalizations supporting \textit{selection}---the most critical atomic operation for interaction---by touch through capacitive sensing and computational inference.  Our computational design pipeline introduces a new design paradigm by concurrently considering the form and interactivity of a physicalization into one cohesive fabrication workflow. We evaluate our approach using (i) computational evaluations, (ii) three usage scenarios focusing on general visualization tasks, and (iii) expert interviews. The design paradigm introduced by our pipeline can lower barriers to physicalization research, creation, and adoption.
}


\graphicspath{{figs/}{figures/}{pictures/}{images/}{./}} 

\usepackage{tabu}
\usepackage{booktabs}     
\usepackage{lipsum}            
\usepackage{mwe}            
\usepackage{mathptmx}  

\usepackage{array}
\usepackage{color,soul}
\usepackage[font={scriptsize,sf},tableposition=top]{caption}
\usepackage{float}
\floatstyle{plaintop}
\restylefloat{table}
\usepackage{makecell}
\usepackage{multirow}
\usepackage{multicol}
\usepackage{colortbl}
\usepackage{pifont}
\usepackage{tabularx}
\usepackage{amsmath}
\usepackage[para,online,flushleft]{threeparttable}
\usepackage{cite,etoolbox}
\usepackage{siunitx}
\usepackage{balance}

\usepackage{amssymb}
\DeclareMathOperator*{\argmax}{arg\,max}

\usepackage{algorithm}
\usepackage[noend]{algcompatible}
\usepackage{setspace}
\floatname{algorithm}{\scriptsize{Algorithm}}

\usepackage[mathcal]{euscript} 
\definecolor{commentout}{rgb}{0.5, 0.5, 0.5}

\begin{document}

\firstsection{Introduction}
\maketitle

\urlstyle{rm}

\setlength{\abovedisplayskip}{3pt}
\setlength{\belowdisplayskip}{3pt}

Interactions are critical for working with data. They help people build mental models as they explore a dataset by zooming into critical details, filtering for important information, or panning to view otherwise hidden surfaces~\cite{liu2010mental, yi2007towards}. Manipulating data physically can further enhance our data understanding when compared to more traditional formats (e.g., 2D display) by being more natural~\cite{drogemuller2021haptic2Dvis}, yielding better performance on certain tasks~\cite{jansen2013evaluating}, and even improving information recall~\cite{stusak2015memorability, stusak2016grasp, kelly2011haptic}.
Despite these benefits, most physicalizations are static representations that do not support common data interactions (e.g., selecting, filtering). Without interactions, physicalizations are essentially static images of datasets, reducing data expressiveness and inhibiting data exploration. 

Creating interactive physicalizations is challenging because it requires cross-disciplinary knowledge spanning visualization, fabrication, and electronics~\cite{bae2022making, djavaherpour2021renderingsurvey}. Visualization enables us to effectively represent data. Fabrication converts data into physical objects using computer-aided designs (CAD). Electronics integrate interactive capabilities into the resulting objects. Most interactive physicalizations are produced in an ad hoc, unstructured manner, resulting in one-off solutions tailored to a specific dataset, problem, or interaction. Without generalizable approaches, these unstructured explorations mean that the form (i.e., physical structure) and function (i.e., interactive capabilities) of physicalizations are produced as separate workflows, leading to a lack of transferable techniques for future physicalizations.

To support generalizability, this work presents a computational design pipeline that enables developers to 3D print various network physicalizations with integrated sensing capabilities.
Given a network dataset, our pipeline leverages multi-material printing to produce network physicalizations that support node \textit{selection} using capacitive sensing---a common technique for capturing touch input on devices (e.g., smartphones)~\cite{grosse2017findingcommon}. The capacitive sensors within the network are then used to uniquely identify each node. By concurrently considering form and interactivity, our pipeline automates low-level hardware instrumentation to raise selection-driven interactions to the application level (e.g., AR/VR, shape-changing displays, or desktop visualizations). Our pipeline intentionally provides flexibility for interaction design: the node selection provides input, but how the selection is used to generate corresponding output on the application level is left to the  developer.

We focus on networks as they have broad utility in almost all domains~\cite{barabasi2016network,newman2018networks,kerren2014multivariate}. Past network physicalizations investigated layouts of physical networks (e.g., brain networks)~\cite{dehmamy2018structural} and how physicality can aid in data sensemaking~\cite{drogemuller2021haptic2Dvis} and common network tasks~\cite{mcguffin2023path}. However, these artifacts lack interactivity. Networks have complex geometry that requires significant engineering considerations to make them physically interactive (e.g., circuit design, sensor integration).
Our computational design pipeline---enabled by 3D printing---addresses these challenges by generating electrical circuits and 3D models in two steps.

First, given a network dataset, our pipeline automatically generates conductive traces (i.e., wires) with tuned electrical resistance that will be integrated into the links of a network physicalization. Second, it generates the necessary 3D geometry to fabricate the sensing network physicalization consisting of links and nodes. Conductive traces within the links act as resistors, and the nodes act as electrodes. The nodes and links are computationally designed to exploit a phenomenon called \textit{resistor-capacitor (RC) delay} (cf.~\autoref{sec3:basic_principles}). Our pipeline generates unique RC time delays when a user touches any node within the network thereby enabling applications to identify the touched node using a single capacitance measure. Applications can use this identification to implement \emph{selection} by touch. Selection---the ability to indicate a mark of interest---is a fundamental interaction primitive for more complex interaction designs, such as filtering or elaborating~\cite{foley1984humanfactors, amar2005lowlevel, yi2007towards}. 

Sensing physicalizations, as a whole, provide a critical step toward creating future physicalizations with interactivity. To lay this foundation, we demonstrate the efficacy of our approach with computational evaluations, expert discussion, and three usage scenarios. These usage scenarios motivate the need for sensing networks and illustrate how selection can support three general visualizations tasks: exploration, explanation, and analytic provenance~\cite{fujiwara2018concise,ynnerman2018exploranation,ragan2016chacterizingprovenance}. Furthermore, our network physicalizations serve as groundwork to support any visualizations modeled as a hub-and-strut form: a series of connected points (e.g., vertices in a Voronoi diagram, points on a line chart).

This work shows how a systematic approach can lead to generalizable techniques for physicalizations by integrating form and interaction into one cohesive fabrication workflow.
Consequently, this new design paradigm can aid in lowering the barrier to physicalization research, creation, and adoption. The source code and supplemental materials (e.g., video demo) are available at our project website~\cite{baesupplmaterials}.

\vspace{3pt}
\noindent \textbf{Contributions:} The primary contributions of this paper are:
\begin{itemize}[topsep=0pt, parsep=0pt, leftmargin=10pt]
    \item A fabrication approach that integrates capacitive sensing into network physicalizations and supports the data interaction \textit{selection};
    \item Novel algorithms that (i) convert network datasets into electrical circuits of resistors, (ii) optimize the resistance selection, and (iii) perform network layout adjustments to satisfy fabrication constraints.
\end{itemize}
\section{Related Work}
Due to the cross-disciplinary nature of physicalizations~\cite{bae2022making}, we build upon past work in (i) interaction primitives in visualization, (ii) fabrication approaches used in physicalizations, (iii) network physicalizations, and (iv) fabrication techniques to create capacitive sensors within 3D printed objects. 

\subsection{Select as an Interaction Primitive}
\label{sec2:select}

To efficiently explore data (whether digitally or physically), people must be able to easily select different parts of the representation and inspect (or hide) details~\cite{ liu2010mental, yi2007towards}. 
Both human-computer interaction (HCI) and visualization recognize \textit{select} as an interaction primitive that enables other interactions to take place. Foley et al.~\cite{foley1984humanfactors} place selection at the highest hierarchical level of input interactions for graphics, indicating its foundational role in supporting more complex interaction designs. Similarly, several visualization interaction taxonomies~\cite{amar2005lowlevel, yi2007towards} highlight the necessity of select for visualization systems to enable other data-related interactions. While select is traditionally achieved with the keyboard and mouse in desktop visualizations, we do not have a corresponding paradigm for physicalizations.  

Despite the lack of interaction standards for physicalization, research identifies several benefits of interacting with data physically. Compared to desktop visualizations, direct manipulation with physicalizations has been shown to be more natural and preferred~\cite{drogemuller2021haptic2Dvis}, yield better task performance~\cite{jansen2013evaluating}, and improve information recall~\cite{stusak2015memorability, stusak2016grasp, kelly2011haptic}. However, most physicalizations still have limited interaction capabilities. A key reason for this limitation is that the materials used to create physicalizations generally do not sense and respond to users.  

Though some physicalizations' forms naturally afford interactions~\cite{Daneshzand2023, Hurtienne2020}, most offset this difficulty by incorporating off-the-shelf electronic components (e.g., motors, LEDs) into their form~\cite{taher2016emerge, legoc2016zooids} or using computer vision (CV) techniques~\cite{herman2021multitouch, mcguffin2023path}. Electronic components have pre-defined scales and dimensions, ultimately affecting how they can be incorporated into a physicalization. Solutions leveraging electronic components do not generalize well to more complex physicalizations (e.g., complex spatial structures, large datasets). For example, electronic components are unlikely able to support interactivity for a network of 50 nodes and 70 links given the complex internal wiring. CV typically relies on visual tracking for gesture interactions, introducing challenges such as inaccurate alignment of the virtual and physical objects as objects move, occlusion from hands or other parts of the structure, and requiring users to stay in-frame~\cite{billinghurst2015survey}. Additionally, most devices using CV cannot reproduce the haptic benefits that we naturally leverage (i.e., holding, rotating, tracing) with our sense of touch. Past studies~\cite{whitlock2020immersive, huang2016embodied, besanccon2017mouse, drogemuller2021haptic2Dvis} confirm the importance of tangible inputs when virtually exploring data. The existing practices for implementing these two approaches reflect the prevailing design paradigm of separating interactive physicalization design into two phases: first thinking about form and then interactions.

Separating form from interaction leads to a series of issues, including post-hoc instrumentation, the potential for conflicting constraints requiring costly design iterations, and a lack of generalizability across design instances. Most existing interaction infrastructures are designed for a specific implementation, resulting in one-off artifacts where the design insights and methods gained from a physicalization are specific to that instantiation~\cite{djavaherpour2021renderingsurvey,bae2022making}. Our methodology integrates sensing capabilities directly into a physicalization's form, enabling designers to concurrently consider form and interactivity. This concurrent design enables us to address two key design and fabrication challenges in physicalization as outlined by Djavaherpour et al~\cite{djavaherpour2021renderingsurvey}: (i) designing for manufacture and assembly and (ii) prototyping and interactive design.

\subsection{Digital Fabrication Approaches for Physicalizations}
\label{sec2:fab}
Digital fabrication uses a digital representation (e.g., 3D model) to produce a physical object. This procedure allows designers to use computational techniques to generate designs for physicalizations that would be otherwise difficult to manually create. 3D printing is a common digital fabrication technique used to make physicalizations~\cite{djavaherpour2021renderingsurvey, bae2022making}. With 3D printing, thin amounts of material are layered to produce an object with complex geometries (e.g., overhangs, lattices, and internal structures). 3D printing can also employ multiple materials to achieve differences in color or material properties (conductivity, flexibility, etc.)~\cite{ngo2018additive}. These capabilities offer creative opportunities to represent data, including cartographic maps~\cite{Allahverdi2018landscaper, thrun2016visualization}, customized data representations~\cite{stusak2014activity}, and common visualization idioms (e.g., bar charts~\cite{swaminathan2014makervis}, networks~\cite{drogemuller2021haptic2Dvis, mcguffin2023path, dehmamy2018structural}). These works illustrate how 3D printing is a powerful representational medium and highlight its potential to produce more complex physicalizations. 

Computational pipelines~\cite{swaminathan2014makervis,abreu2022pipeline} can also make designing physicalizations easier. MakerVis~\cite{swaminathan2014makervis} imports a dataset to be fabricated as layered visualizations (e.g., bar, line) or prism maps. These physicalizations are passive, static objects. Abreu de Freitas et al.~\cite{abreu2022pipeline} extend MakerVis by enabling a designer to specify the intended behavior for bar charts (i.e., passive, reconfigurable, dynamic). Dynamic bars are printed but must be attached to various external electronic components for interaction, which presents limitations (cf. \autoref{sec2:select}). 

We build on past pipelines in two ways: supporting structural complexity using 3D printing and integrating key interaction support during fabrication. Our approach explores complex geometries (i.e., networks), which offer unique structural challenges for fabrication that our pipeline dynamically resolves (see \autoref{sec3:network-layout}). Our pipeline also automatically specifies and produces electrical circuitry inside of a physicalization using a conductive filament, rather than designing around electrical components that would be integrated post-production. This approach eases the process of creating interactive physicalizations by linking electronic design and implementation with a physicalization's form.

\subsection{Network Physicalizations}
Current network physicalizations are passive objects with no innate sensing capabilities (e.g.,~\cite{freksa2019geometric, grabkowska2019network, stewart2019network}). Drogemuller et al.~\cite{drogemuller2021haptic2Dvis} investigate haptic and visual comprehension of a flat 2D network physicalization (16--24 nodes) using an external camera to log user interactions. Dehmamy et al.~\cite{dehmamy2018structural} introduce a network layout algorithm for physical networks that is optimized to avoid link intersection, but their algorithm relies on a rigorous trial-and-error process to determine layout parameters. Their 3D printed network (184 nodes, 176 links) does not support integrated sensing.
McGuffin et al.~\cite{mcguffin2023path} show the promise of physical network interactions (70 nodes, 140 links) mediated with augmented reality (AR) tracked using external cameras, which present interaction limitations (cf. \autoref{sec2:select}). In contrast to these works, our approach supports network physicalizations with minimal parameter-tuning to produce free-standing networks with integrated sensing capabilities.

\subsection{Integrating Interactivity into 3D Printed Objects}
Most objects made on 3D printers are static forms that have no interactive capabilities. HCI research explored embedding electronic components (e.g., sensors and LEDs~\cite{he2022modelec, willis2012printedOptics, weller2008posey}) and designing internal structures (e.g., pipes for light~\cite{savage2014pipes, willis2012printedOptics}) in printed objects to support interactivity through user input and display output. Past work also investigated how to create conductive traces within printed objects to achieve \emph{capacitive sensing}~\cite{burstyn2015printput, schmitz2015capricate,schmitz2019trilaterate}, a common technique used in commercial devices (e.g., smartphones, tablets) to capture touch input~\cite{grosse2017findingcommon}. A conductive material (e.g., conductive PLA) inside an object can create electrical traces for capacitive sensing. Capricate~\cite{schmitz2015capricate} and ./trilaterate~\cite{schmitz2019trilaterate} create capacitive sensors in 3D printed forms; however, both rely on having an individual electrical connection (i.e., wire) for each sensor. This instrumentation can affect an object's mobility and scalability. Furthermore, neither system is designed to handle the complex structure of networks. Our approach requires a minimum of one wire to achieve sensing and optimizes the design of electrical traces, capacitive sensors, and physical geometry specifically for networks.  

\section{Methodology}
\label{sec3:metholodogy}
\begin{figure}[tb]
	\centering
	\includegraphics[width=\linewidth]{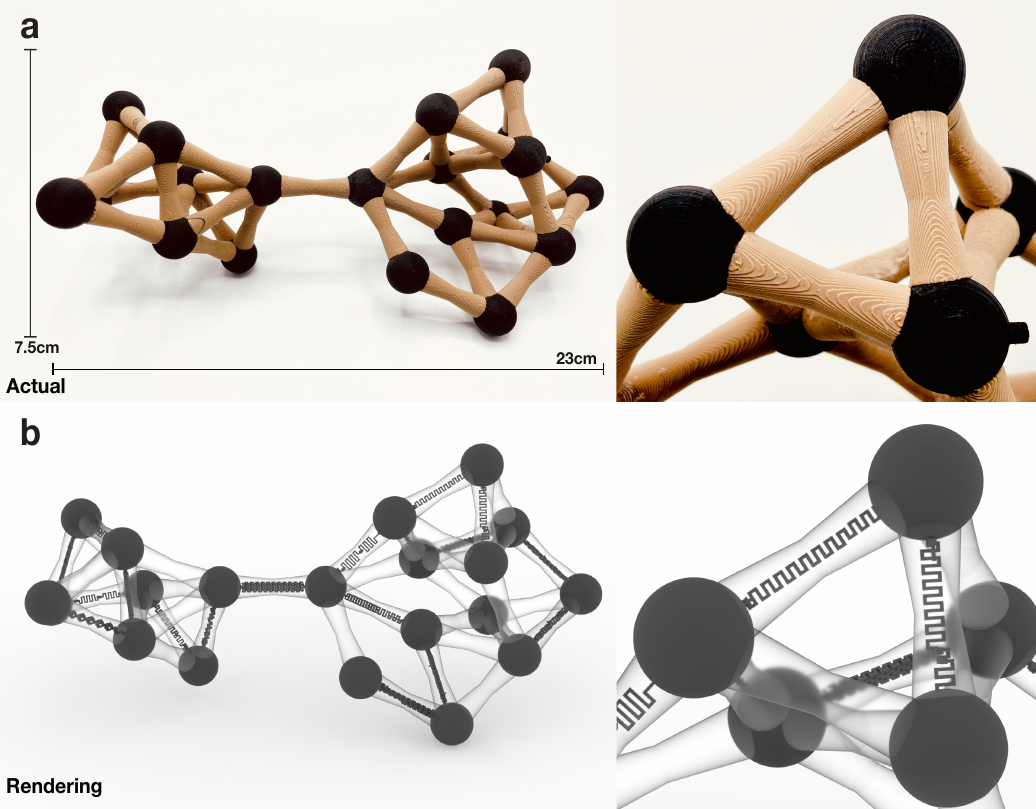}
    \caption{A sensing network physicalization ($N=20$, $L=40$). (a) A multi-material 3D printed network physicalization produced by our computational design pipeline (\autoref{sec3:overview}). Conductive traces are embedded in the network's links which enables node selection via capacitive sensing. (b) A computational rendering of the network physicalization showcasing how the conductive traces are distributed throughout the network's links. The conductive traces use a serpentine pattern.}  
    \label{fig:sec3:network-overview}
    \vspace{-1em}
\end{figure}

Our computationazl design pipeline produces network physicalizations capable of node selection via capacitive sensing (\autoref{fig:sec3:network-overview}a). 
When a user touches a node, the network physicalization outputs unique sensor readouts such that a visualization application (e.g., desktop, AR/VR) can identify the selected node. See \autoref{sec4:usage-scenarios} for examples of how these networks support a range of analytical scenarios.
We first outline the basic principles of how we achieve capacitive sensing using multi-material 3D printing (\autoref{sec3:basic_principles}). We then provide an overview of the pipeline (\autoref{sec3:overview}) and discuss each pipeline subcomponent in \autoref{sec3:circuit-design}--\autoref{sec3:wiring}.
The pipeline involves various network representations. To distinguish these networks, we define and use the following terms: 
{\begin{itemize}[topsep=0pt, parsep=-1pt, leftmargin=10pt]
  \item \textbf{Network Dataset:} Dataset pertaining to nodes and links.
    \item \textbf{Resistor Network:} A network of resistors derived from the network dataset that will be converted into conductive traces. 
    \item \textbf{3D Representation Network:} A 3D representation of the network dataset, assigning spatial positions to nodes and links.
    \item \textbf{Fabrication-ready Network:} A digital model for 3D printing that embeds the resistor network within the 3D representation network
    \item \textbf{Printed Network:} A 3D printed network produced from the fabrication-ready network 3D model
    \item \textbf{Sensing Network:} A printed network that has sensing capabilities.
\end{itemize}}
\noindent Our supplementary materials summarize notations used in the paper~\cite{baesupplmaterials}.

\subsection{Basic Principles of Sensing Network Physicalization}
\label{sec3:basic_principles}
\begin{figure}[tb]
	\centering
	\includegraphics[width=\columnwidth]{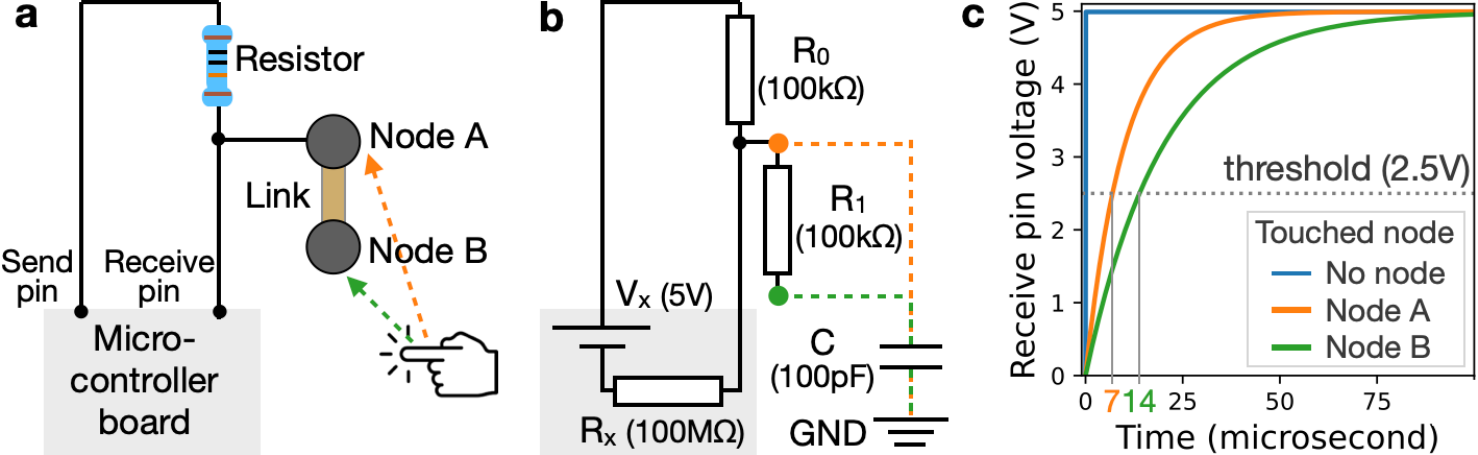}
    \caption{Identifying node selection with capacitive sensing: (a) overall schematic where a two-node printed network is connected to a microcontroller's circuit; (b) the circuit diagram corresponding to (a) with representative resistance measurements; and (c) the voltage change measured at the receive pin when a different node is touched. 
    }
    \label{fig:basic_principle}
    \vspace{-0.5em}
\end{figure}

We use capacitive sensing to infer node selection by touch.
This capacitive sensing is achieved through the conductive traces distributed throughout a set of a network's links (\autoref{fig:sec3:network-overview}b).
Conductive filament used in 3D printing can act as resistors in electrical circuits~\cite{dijkshoorn2019characterizing}. Our printed network has (i) conductive nodes and (ii) traces of specific electrical resistance integrated into its links. As shown in \autoref{fig:basic_principle}a (a two-node network), we connect this printed network to an electrical circuit that is managed by a microcontroller. This connection results in a combined circuit as shown in \autoref{fig:basic_principle}b.

When a user touches a node (e.g., in \autoref{fig:basic_principle}b, the orange and green points are nodes A and B, respectively), the user's body and the node become capacitively coupled (i.e., energy moves between them~\cite{grosse2017findingcommon}), inducing an \textit{RC delay}. RC delay is the time required to charge a capacitor in a circuit through a particular amount of resistance. Increasing the resistance in the circuit will also increase the amount of time needed to charge the capacitor. Our pipeline is based on this principle: we can design networks where the RC delay varies based on which node is touched (cf. \autoref{fig:basic_principle}c).
By computationally designing the geometry of conductive traces, different paths with unique amounts of resistance and regions can function as entry points of capacitive coupling for the RC delay. 

Thus, each capacitive region takes a different amount of time to charge once a user touches it. 
This process enables us to infer a touched node by measuring the time taken to reach a predefined voltage threshold (e.g., 2.5V) on a microcontroller. For example, 0\si{\micro\second} indicates no nodes are touched; 7\si{\micro\second} is Node A, and 14\si{\micro\second} is Node B. 
Our pipeline automatically converts network data into an electrical circuit that can uniquely identify each node with capacitive sensing with the following.

\textbf{Conductive printed network.} For a printed network, nodes should have high conductivity, so their resistance is essentially negligible. Links, in contrast, should have low conductivity so they can act as resistors. This process requires converting network data, specifically the links, into a network of resistors as \textit{conductive traces} (i.e., electrical paths). The resistance of these traces can be tuned by varying their length and thickness~\cite{dijkshoorn2019characterizing}. In our case, low conductivity/high resistance can be achieved by printing a long, thin line using a conductive filament with specific geometry computed using the resistivity law:
\begin{equation}
\label{eq:resistivity-law}
    r = \frac{\rho l}{a}
\end{equation}
where $r$ is the resultant resistance, $\rho$ is a material's resistivity, and $l$ and $a$ are respectively the length and cross-sectional area of an object. Nodes are printed using a conductive filament, producing a large cross-sectional area per length, resulting in low resistance. Links enclose conductive traces using non-conductive filament, which can be varied to support different visual design parameters (e.g., length, width, color).

\begin{figure*}[tb]
	\centering
	\includegraphics[width=\textwidth]{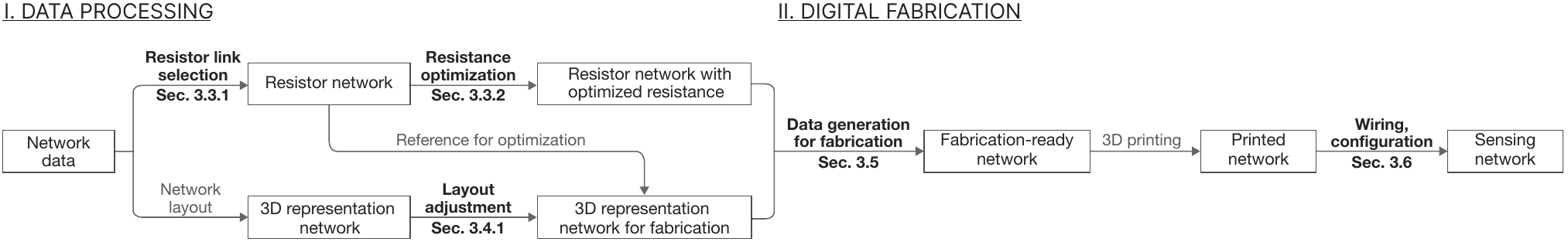}
    \caption{A computational design pipeline for producing sensing network physicalizations. Rectangles represent output data or physical objects (see \autoref{sec3:metholodogy} for definitions). Arrows represent different processes. Processes that are bolded are new methods that we introduce (\autoref{sec3:circuit-design}--\autoref{sec3:wiring}).}  
    \label{fig:sec3-pipeline}
    \vspace{-1em}
\end{figure*}

\textbf{Capacitive sensing of different nodes.}
When an electrical circuit consists of a power source, a resistor, and a capacitor in series, the voltage change of the capacitor is expressed as: 
\begin{equation}
\label{eq:voltage-transient}
    v(t) = v_{\mathrm{in}} \left(1 - e^{-\frac{t}{c r}}\right)
\end{equation}
where $t$ is the time from the start of charging the capacitor; $v_{\mathrm{in}}$ is the input voltage from the power source; $c$ is the capacitor's capacitance; and $r$ is the resistor's resistance. 
When we apply capacitive sensing to a printed network, the equation changes based on the number of links (i.e., resistors) and their resistance, the resistors' connections (i.e., series, parallel, or combination),
and the touched node (i.e., a connection between a user/node). 

The pipeline transforms a network dataset into a resistor network, which is optimized to uniquely identify nodes through capacitive sensing. 
Then our pipeline automatically generates the necessary CAD geometry (i.e., fabrication-ready network) that can be exported as STL files for 3D printing.
After 3D printing, we can directly produce a sensing network physicalization by connecting the printed network to an electrical circuit that is managed by a microcontroller (\autoref{fig:basic_principle}).

\subsection{Computational Design Pipeline Overview}
\label{sec3:overview}
Our computational pipeline (\autoref{fig:sec3-pipeline}) is divided into two stages: data processing (\autoref{sec3:circuit-design}, \autoref{sec3:network-layout-overview}) and digital fabrication (\autoref{sec3:digital-fab}, \autoref{sec3:wiring}). 

\textbf{Data processing.}
Data processing has two parallel tracks. The first track (top) designs the resistor network and the second track (bottom) adjusts the network layout to satisfy fabrication constraints. The resistor network design begins by selecting links from the network dataset to be used as resistors. We refer to the selected links as \textit{resistor links}. Then our resistance optimization process identifies a suitable resistance for each resistor link such that the RC delay is different for every touched node. The network layout design begins by laying out the network dataset as a 3D representation network. For this process, we can use any existing network layout algorithms or use predefined positions specified by the source network data. Next, we apply a network layout adjustment method using neural networks. This adjustment is primarily to prevent conductive components (i.e., nodes and resistor links) from intersecting to avoid interference with the circuit (\autoref{fig:layout_adjustment}). 

\textbf{Digital Fabrication.} The digital fabrication stage uses the outputs from the data processing stage to dynamically generate a fabrication-ready network (i.e., CAD files). 
This process involves computationally (i) drawing the resistor network as long, thin traces and (ii) updating the link geometry to structurally reinforce the connection between nodes and links.
We fabricate our network using multi-material printing and then connect the printed network to a microcontroller. 

\subsection{Resistor Network Design}
\label{sec3:circuit-design}
Designing the resistor network consists of selecting a subset of links to be used as resistor links and optimizing their resistance to achieve unique RC delays when a node is touched. 
Note that we only consider networks with a single connected component (i.e., no nodes are completely isolated from others).

\subsubsection{Resistor Link Selection}
\label{sec3:resitor-selection}
Networks typically contain more links than nodes (e.g., 10 nodes, 30 links), but, in our approach, all links do not need to be resistors to electrically connect all nodes for sensing. See \autoref{fig:sec3:network-overview}b. For a network with $N$ nodes and $L$ links, we only need to find $(N-1)$ resistor links that connect all $N$ nodes. The minimum set of such links can be found by performing a traversal search such as breadth-first search (BFS) or depth-first search (DFS) over the network. 

\begin{figure}[tb]
	\centering
	\includegraphics[width=0.93\columnwidth]{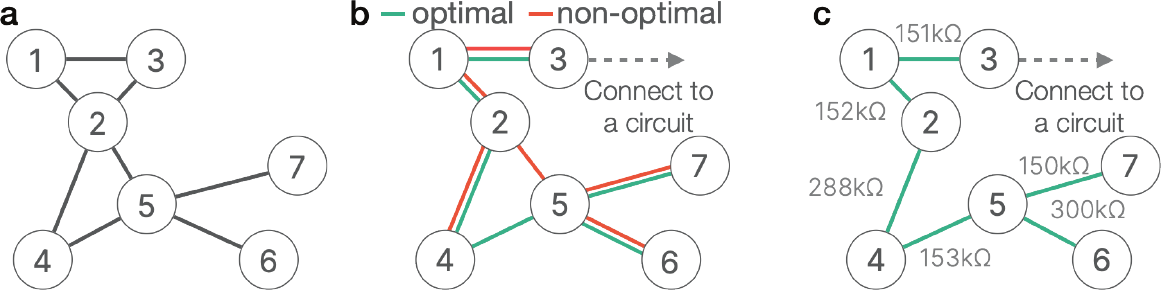}
    \caption{Example of resistor network generation: (a) input network, (b) resistor network with optimal resistor links (green), while red links are non-optimal, (c) resistor network after assigning optimized resistance.}
    \label{fig:circuit_network_design}
\end{figure}

We use DFS to reduce the computational complexity of the resistance optimization process in \autoref{sec3:resistance_opt}. DFS selects a set of resistor links that generate fewer path branches compared to BFS. However, based on the starting node of the traversal, DFS may find a set of resistor links with more path branches. For example, in \autoref{fig:circuit_network_design}b, we show two different sets of resistor links selected from the network dataset in \autoref{fig:circuit_network_design}a.
The green resistor links have the minimum number of branches (two branches: nodes 3--6 and 5--7). The red resistor links have three branches (nodes 3--6, 2--4, and 5--7).
To find resistor links with a fewer number of path branches, we run DFS $N$ times. 
Each iteration selects a different node as a starting node, and then we select the result with the minimum number of branches. Note that we use this DFS-based selection as a heuristic because finding the optimal solution (i.e., the fewest number of path branches) can be computationally expensive~\cite{furer1992approximating}. After determining the resistor links, we also select one or two leaf nodes as the connection point(s) to the microcontroller's circuit (e.g., Node 3 in \autoref{fig:circuit_network_design}). Then, for each resistor link, we calculate its appropriate resistance using the resistance optimization (cf. \autoref{fig:circuit_network_design}c).

\subsubsection{Resistance Optimization}
\label{sec3:resistance_opt}

\textbf{Optimization goal.}
We can identify which node is touched based on how long it takes for a capacitor to reach a certain voltage threshold (i.e., RC delay). When a user touches a node, a new connection to a capacitor (i.e., the user's body) and the ground is introduced into the electrical circuit. This connection varies based on the touched node (e.g., the orange vs. green dashed wires in \autoref{fig:basic_principle}b), inducing a different RC delay based on the circuit design. Although slight time delays are theoretically sufficient to recognize the selection of different nodes (e.g., \SI{10}{\nano s}), we want to \textbf{maximize the minimum difference among the time delays} for fault tolerance in the microcontroller and 3D printer. First, we assume a microcontroller is used to capture time delays. The CPU clock of microcontrollers is typically slow (e.g., \si{16\mega\hertz} for the Arduino UNO R3~\cite{arduinouno}), which limits the precision of the measurement (e.g., \si{62\nano\second} for \si{16\mega\hertz} CPUs). Second, errors can be introduced during the 3D printing process.
Our optimization goal is formulated as:
\begin{equation}
\label{eq:resistance_opt}
    \argmax_{\substack{\mathbf{r} \in \mathbb{R}^{N-1}, \, r_\mathrm{min} \leq r_k \leq r_\mathrm{max}}} \min\{d_{i,j}(\mathbf{r}) | 1 \leq i \leq N, 1 \leq j \leq N \}
\end{equation}
where $\mathbf{r}$ is a vector of resistance assigned to $(N-1)$ resistor links, $r_\mathrm{min}$ and $r_\mathrm{max}$ specify the value boundary for each resistance $r_k$ in $\mathbf{r}$, and $d_{i,j}(\mathbf{r})$ is the difference of the time delays between cases when nodes $i$ and $j$ are touched.
We define the difference of the time delays as $d_{i,j}(\mathbf{r}) = \| t_i(\mathbf{r}) - t_j(\mathbf{r}) \|$, where $t_i(\mathbf{r})$ is the time delay when node $i$ is touched.
We call $t_i$ node $i$'s \textit{time delay function}.

The optimization for \autoref{eq:resistance_opt} has two challenges.
First, the time delay function differs based on the selected node as touching a new node results in a different circuit.
Consequently, we need to evaluate various different circuits.
Second, we need to optimize $\mathbf{r}$, which has a large number of parameters (i.e., $N{-}1$). 
To address the first challenge, we leverage circuit simulators to computationally generate and evaluate a circuit corresponding to each touched node.
For the second, we compute the derivative of each time delay function with respect to $\mathbf{r}$ and then apply gradient descent~\cite{ruder2016overview} to efficiently optimize all parameters.

\textbf{Circuit generation and simplification.}
The pseudocode for our resistance optimization algorithm is shown in \autoref{alg:resistance-opt}.
The optimization starts by registering the circuit information (including the microcontroller's circuit and resistor links) with a circuit simulator (line 1). Afterward, we generate $N$ different circuits, each of which corresponds to a case where one of $N$ nodes is touched (line 3). Circuit simulators (e.g., SPICE~\cite{tuinenga1995spice}) typically have an analysis function to observe voltage changes. We thus avoid manually deriving all necessary equations and functions to calculate the time delay caused by each touch.

Among a variety of circuit simulators, we employ a simulator (e.g., Lcapy~\cite{hayes2022lcapy}) that accepts symbolic computations~\cite{meurer2017sympy,cohen2003computer} to derive derivatives.
In symbolic computations, we can use mathematical symbols as-is to solve equations and compute derivatives without converting them to numerical values.
Consequently, we can obtain precise derivatives for the optimization and also perform an efficient optimization by reusing the same symbolic functions across iterations. 

\begin{algorithm}[tb]
    \setstretch{0.83}
    \caption{\scriptsize{Resistance optimization.}}
    \label{alg:resistance-opt}
    \small
    \algblockdefx{FORALLP}{ENDFORALLP}[1]%
  {\textbf{for all }#1 \textbf{do in parallel}}%
  {\textbf{end for}}
\algtext*{ENDFORALLP}

\textbf{Inputs:} $\mathcal{N}$, network nodes; $\mathcal{R}$, resistor links; $\mathcal{M}$, microcontroller circuit information; C, artificial human capacitor; v$_\mathrm{thres}$, the voltage threshold to measure the time delay; $[r_\mathrm{min}, r_\mathrm{max}]$, boundary of possible resistance
\\
\textbf{Outputs:} $\mathbf{r}_\mathrm{best}$, a set of optimized resistance

\begin{algorithmic}[1]
\STATEx \underline{Prepare touched circuits and corresponding symbolic functions}
\STATE Register $\mathcal{M}$ and $\mathcal{R}$ with a symbolic circuit simulator.
\FORALLP{$\mathcal{N}$}
  \STATE Create a touched circuit $\mathcal{M}'_i$ by adding C into $\mathcal{M}$.
  \STATE Combine $\mathcal{M}'_i$'s resistors connected in series.
  \STATE Get a mapping between resistance before ($\mathbf{r}$) and after ($\mathbf{r}'_i$) aggregation.
  \STATE Generate a time-varying voltage function $v_i$ at the receive pin.
  \STATE Obtain a symbolic time delay function $t_i$ by solving v$_\mathrm{thres} = v_i(t_i)$.
\ENDFORALLP
\FOR{$i = 1, \cdots, |\mathcal{N}|$}
    \FOR{$j = i + 1, \cdots, |\mathcal{N}|$}
        \STATE Compute the symbolic time delay difference function $d_{i,j} = \| t_i - t_j \|$.
        \STATE Compute $\nabla d_{i,j}$, the symbolic derivative of $d_{i,j}$ with respect to $\{\mathbf{r}'_i, \mathbf{r}'_j\}$.
    \ENDFOR
\ENDFOR

\STATEx \underline{Iterative optimization using gradient descent}
\STATE Initialize $\mathbf{r}$ (e.g., assigning random values in a range of $[r_\mathrm{min}, r_\mathrm{max}]$).
\WHILE{not converged}
    \STATE Evaluate \autoref{eq:resistance_opt} with current $\mathbf{r}$ and keep the best result so far as $\mathbf{r_\mathrm{best}}$.
    \STATE Find a bottleneck pair of nodes ($x$, $y$) that produced the minimum $d_{i,j}$.
    \STATE Update $\mathbf{r}$ based on $\nabla d_{x,y}(\{ \mathbf{r}'_x, \mathbf{r}'_y\})$ and the mapping between $\mathbf{r}'_i$ and $\mathbf{r}$.
    \STATE Clip $\mathbf{r}$ to fit resistance into the boundary $[r_\mathrm{min}, r_\mathrm{max}]$.
\ENDWHILE

\STATE \textbf{return} $\mathbf{r_\mathrm{best}}$

\end{algorithmic}
\end{algorithm}

Symbolic computing still suffers from computational costs when several symbols are involved, particularly when we compute the matrix inversion operation required for solving our optimization equations. The time complexity for this matrix inversion operation is $\smash{\mathcal{O}(S^3)}$ where $S$ is the number of symbols. To mitigate this complexity, we utilize the fact that resistors connected in series can be simplified as one resistor~\cite{brown2003series}. Given three resistors  (i.e., $r_1$, $r_2$, $r_3$), the symbolic computation without combining these resistors would involve three symbols. In contrast, combining three resistors with $r' = r_1 + r_2 + r_3$ would only involve $r'$ in the computation.

Combining resistors can radically reduce computational costs.
To maximally reduce the number of symbols, we use the DFS-based resistor link selection as discussed in \autoref{sec3:resitor-selection}. We combine resistors for each of the $N$ generated circuits, derive their voltage change functions, and solve each function for a given voltage threshold (lines 4--7 in \autoref{alg:resistance-opt}). We further speed up these processes by solving the functions in parallel.
For the circuit induced by touching node $i$,  we denote a vector of resistance of the combined resistors as $\mathbf{r}'_i$, a bijective mapping from $\mathbf{r}$ to $\mathbf{r}'_i$ as $f_i: \mathbf{r} \rightarrow \mathbf{r}'_i$ and its inverse mapping as $\smash{f_i^{-1}: \mathbf{r}'_i \rightarrow \mathbf{r}}$. Note: $|\mathbf{r}'_i| \ll |\mathbf{r}|$.

\textbf{Gradient-based optimization.}
From all pairs of the time delay functions, we symbolically derive the time delay difference (i.e., $d_{i,j}$) and its derivative ($\nabla d_{i,j}$) with respect to $\{\mathbf{r}'_i, \mathbf{r}'_j\}$ (lines 8--11).
We then assign concrete values to $\mathbf{r}$ and iteratively optimize these values. 

We first initialize $\mathbf{r}$ with user-specified or random values within a range of $[r_\mathrm{min}, r_\mathrm{max}]$ (line 12).
We evaluate \autoref{eq:resistance_opt} with the current $\mathbf{r}$ (line 14) by inserting the values of $\mathbf{r}$ into $d_{i,j}$ while referring to the mappings $f_i$ and $f_j$.
We then update $\mathbf{r}$ to improve the minimum time delay difference with gradient descent.
To achieve this, we first identify the bottleneck $d_{x,y}$, causing the minimum time delay difference (line 15).
We then update $\mathbf{r}$ based on $\nabla d_{x,y}(\{\mathbf{r}'_x, \mathbf{r}'_y\})$.
However, the obtained gradients are for the combined resistors.
Thus, we evenly distribute each combined resistance's gradient to the set of original resistor links by referring to $\smash{f_x^{-1}}$ and $\smash{f_y^{-1}}$.
For example, when a combined resistor consists of $\{r_1, r_2, r_3\}$ and its gradient is $3$, we assign $1$ as a gradient for each of $\{r_1, r_2, r_3\}$.
For convenience, we denote the gradients after this conversion as $\mathrm{grad}(d_{x,y}) (\mathbf{r})$.
We can formulate the update of $\mathbf{r}$ by gradient descent (line 16) 
as:
\begin{equation}
    \mathbf{r} \gets \mathbf{r} - \alpha \, \mathrm{grad}(d_{x,y}) (\mathbf{r})
\end{equation}
where $\alpha \in \mathbb{R}$ is a step size
to control the pace of the optimization (e.g., 1\% of $(r\_\mathrm{max} - r\_\mathrm{min})$).
We then clip each resistance of $\mathbf{r}$ in the user-specified boundary, $[r_\mathrm{min}, r_\mathrm{max}]$ (line 17). 
We repeat until the optimization reaches convergence and then use the value of $\mathbf{r}$ that achieves the best result for \autoref{eq:resistance_opt} (lines 13--18).

\begin{figure}[tb]
	\centering
	\includegraphics[width=\columnwidth]{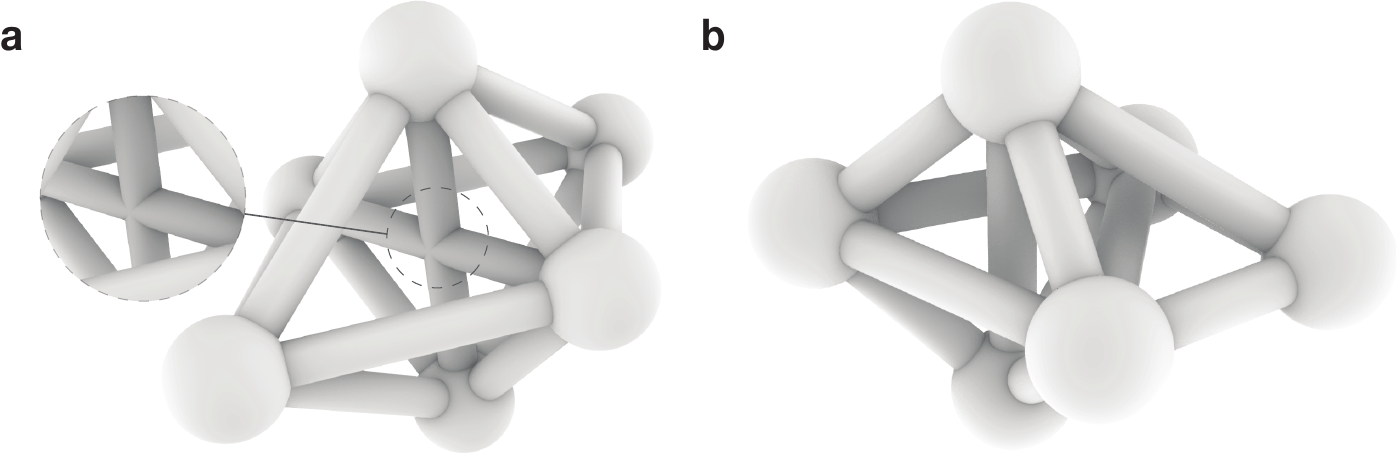}
    \caption{A comparison of network layouts before and after adjustment. 
    Two links intersect, creating potential circuitry issues in (a). Spatial adjustment resolves this issue (b).}
    \label{fig:layout_adjustment}
    \vspace{-0.5em}
\end{figure}

\subsection{Network Layout Design}
\label{sec3:network-layout-overview}
For the network layout design, we generate a 3D representation network. 
Our pipeline conducts post-hoc adjustments that work with any existing network layout algorithm.
By default, we use the force-directed Fruchterman-Reingold layout~\cite{fruchterman1991graph}. 
We can also use predefined positions if they are available (e.g., networks with molecular structures, brain regions, geospatial information). This process extends to non-network visualizations deconstructed as hubs and struts (e.g., vertices in a Voronoi diagram, points in a line graph)  by using the same process as a network with predefined positions. In this work, we mainly consider networks only with straight links. 

\subsubsection{Network Layout Adjustment}
\label{sec3:network-layout}
The 3D representation network generated from the above process might have an undesirable structure for fabrication.
Since most existing layout algorithms do not consider 3D volumes of nodes and links, the conductive nodes and resistor links might intersect with each other, resulting in unexpected changes in the circuit. 
Thus, we introduce a network layout adjustment algorithm to adjust network geometry to avoid such issues.
\autoref{fig:layout_adjustment} shows an example of 3D representation networks before (\autoref{fig:layout_adjustment}a) and after (\autoref{fig:layout_adjustment}b) applying our layout adjustment.

\textbf{Optimization goals.} Our layout adjustment has the following goals: 
\begin{itemize}[topsep=0pt, parsep=0pt, leftmargin=10pt]
    \item Primary goal: All resistor links \textit{must} not intersect with each other. This restriction also applies to all nodes. 
    \item Secondary goal: The adjusted layout should preserve the structure of the original layout as much as possible.
    \item Optional aesthetic goals:
    \begin{itemize}[topsep=0pt, parsep=0pt, leftmargin=10pt]
        \item Intersections of non-conductive links should also be minimized.
        \item Link lengths should be close to uniform~\cite{bennett2007aesthetics}.
    \end{itemize}
\end{itemize}
To efficiently satisfy these multiple goals, we employ a neural network (NN). A NN enables us to conveniently perform gradient-based optimization over loss functions corresponding to these goals~\cite{ahmed2022multicriteria}.
The optional goals do not fulfill all graph aesthetic criteria (e.g., maximizing link angles). However, our NN approach can readily extend to support such options by designing additional corresponding loss functions.

\textbf{Loss functions.} Node-link diagrams are intuitively represented as cylinders (links) and spheres (nodes). We can detect the intersection of two cylinders if the minimum distance of their axes is smaller than the sum of their base radii. Similarly, two spheres intersect if the distance between their centers is smaller than the sum of their radii. Let $\mathrm{dist}(\cdot)$ be a generic function computing the distance of links' cylindrical axes or the distance of nodes' spheres; $\mathrm{radius}(\cdot)$ be a generic function returning a radius of the link's cylinder base or node's sphere.
We define an intersection loss for all links and nodes as:
\begin{equation}
    J_\mathrm{int} = \sum_i \sum_{j, j \neq i} \mathrm{ReLU}\left(\mathrm{dist}(e_i, e_j) - \mathrm{radius}(e_i) - \mathrm{radius}(e_j)\right)
\end{equation}
where $\smash{\{e_i\}}$ is a set of nodes or links and $\smash{\mathrm{ReLU}(\cdot)}$ is the rectified linear unit function ($\smash{\mathrm{ReLU}(x) = \max(0, x)}$).
We use $\mathrm{ReLU}$, instead of thresholding, to train the NN more easily.
If there is no intersection, $\mathrm{ReLU}$ returns 0; otherwise, it returns the intersected length. 
We compute the intersection loss separately for resistor links ($\smash{J_\mathrm{\scriptscriptstyle int}^\mathrm{\scriptscriptstyle res}}$), non-resistor links ($\smash{J_\mathrm{\scriptscriptstyle int}^\mathrm{\scriptscriptstyle non\_res}}$), and nodes ($\smash{J_\mathrm{\scriptscriptstyle int}^\mathrm{\scriptscriptstyle node}}$).

Similar to Wang et al.~\cite{wang2020deepdrawing}, we measure the network layout change before and after adjusting node positions with Procrustes distance. Procrustes distance computes the sum of Euclidean pairwise distances of nodes after aligning two sets of node positions with translation, uniform scaling, rotation, and axis flipping~\cite{fujiwara2020incremental}. This measure can eliminate the influence of non-essential layout differences. Let $\smash{\mathbf{P}^\mathrm{o}}$ and $\smash{\mathbf{P}^\mathrm{u}}$ be the node positions of the 3D representation network before and after the layout adjustment.
Then the loss for the layout change can be written as: $\smash{J_\mathrm{\scriptscriptstyle pos} {=} \mathrm{Procrustes}(\mathbf{P}^\mathrm{o}, \mathbf{P}^\mathrm{u})}$ where $\smash{\mathrm{Procrustes}(\cdot)}$ computes the Procrustes distance. Lastly, we measure a loss corresponding to non-uniform link lengths, $J_\mathrm{\scriptscriptstyle len}$, as the standard deviation of link lengths.

\textbf{NN architecture and optimization procedure.}
We use a NN to find appropriate updated 3D node positions, $\smash{\mathbf{P}^\mathrm{u}}$. Thus, the NN outputs $\smash{\mathbf{P}^\mathrm{u}}$ from $\smash{I_{\scriptscriptstyle N}}$, an input identity matrix with the size of $N$. We also input the node and link information required to compute the loss functions.
We expect most network physicalizations to be relatively small (e.g., less than 1000 nodes). Therefore, we use a simple multilayer perceptron (MLP) consisting of three 100-neuron hidden layers by default.

Our training consists of two phases: (i) learning the original layout and (ii) adjusting the layout.
The first phase learns neuron weights such that $\smash{\mathbf{P}^\mathrm{u}}$ becomes the same layout as $\smash{\mathbf{P}^\mathrm{o}}$ by minimizing $J_\mathrm{\scriptscriptstyle pos}$ during training. 
The second phase then adjusts the learned neuron weights to satisfy our optimization goals. 
For this second phase, we minimize the sum of multiple weighted loss functions. 
To place a higher priority on removing intersections, we assign larger weights to 
$\smash{J_\mathrm{\scriptscriptstyle int}^\mathrm{\scriptscriptstyle res}}$ and $\smash{J_\mathrm{\scriptscriptstyle int}^\mathrm{\scriptscriptstyle node}}$.
For the loss functions corresponding to the secondary goal of layout preservation and optional aesthetic goals, we can select weights based on the design needs.
For example, when generating the network layout shown in \autoref{fig:sec3:network-overview}a, we used a loss function, $3 \smash{J_\mathrm{\scriptscriptstyle int}^\mathrm{\scriptscriptstyle res}} + 3 \smash{J_\mathrm{\scriptscriptstyle int}^\mathrm{\scriptscriptstyle node}} + \smash{J_\mathrm{\scriptscriptstyle int}^\mathrm{\scriptscriptstyle non\_res}} + J_\mathrm{\scriptscriptstyle pos} + J_\mathrm{\scriptscriptstyle len}$. If $\smash{J_\mathrm{\scriptscriptstyle int}^\mathrm{\scriptscriptstyle res}} > 0 $ or  $\smash{J_\mathrm{\scriptscriptstyle int}^\mathrm{\scriptscriptstyle node}} > 0$, the optimization result does not satisfy the primary goal. To resolve such cases, we can retrain the NN with larger weights for $\smash{J_\mathrm{\scriptscriptstyle int}^\mathrm{\scriptscriptstyle res}}$ and $\smash{J_\mathrm{\scriptscriptstyle int}^\mathrm{\scriptscriptstyle node}}$ and/or reduce the links' base radii relative to the node radii.

\subsection{Data Generation for Fabrication}
\label{sec3:digital-fab}
The digital fabrication stage creates the necessary CAD files to 3D print the updated network. We generate a fabrication-ready network by embedding the resistor network into the 3D representation network. 

\textbf{Resistor network embedding.} 
Nodes are fabricated with a conductive filament to provide high conductivity.
To materialize resistor links, they need to have high resistance within a limited volume of each cylindrical link. This is accomplished by drawing a thin, long trace of the conductive filament using a serpentine trace pattern~\cite{soh2009comprehensive}. Given a surface area, the resistivity is maximized by drawing on the $xy$-plane layers with a serpentine trace pattern and connecting the endpoints in the $z$-direction line, resulting in a 3D zig-zag structure (\autoref{fig:sec3:network-overview}b).

To achieve the resistance specified in the resistor network, we determine the length of each resistor link's trace based on the resistivity law (\autoref{eq:resistivity-law}).
Due to the printing resolution of most FDM printers, we suggest using different thicknesses for the line on $xy$-plane and the line along $z$-direction.  
The line thickness for the $xy$-plane can be close to the printer's nozzle extrusion width (e.g., \SI{0.4}{\milli\meter}), while the thickness for $z$-direction should be at least twice the extrusion width (e.g., \SI{0.8}{\milli\meter}) to ensure contact from the previous layer. 
A conductive filament's resistivity (i.e., $\rho$ in \autoref{eq:resistivity-law}) along the $xy$-plane and $z$-direction is typically provided by the filament makers. 
Then, with the given line thickness, filament's resistivity, and link's cylindrical shape (radius, angle, and length), we can computationally identify the appropriate length and drawing pattern of the trace.
However, various external conditions may influence the line area and resistivity (e.g., printing precision, nozzle temperature, and filament production errors). 
Thus, we recommend identifying resistance per length (i.e., $\rho/a$) for the $xy$-plane and $z$-direction under the expected printing condition. 
The supplementary materials include our process used to obtain this information.

\textbf{Structural support.} 
Our sensing network is a physical interface, and thus needs to be structurally sound.
We add cone-shaped structural support to the edges to increase contact.

\subsection{Wiring and Calibration}
\label{sec3:wiring}
We 3D print a fabrication-ready network to produce a printed network. We connect it to a microcontroller circuit using the same schematic diagram as \autoref{fig:basic_principle}a. We calibrate time delays corresponding to all nodes by manually touching each node and observing the time required to reach a microcontroller's logic threshold voltage (e.g., 2.5V in \autoref{fig:basic_principle}c). Calibration is necessary as each individual and external factors (e.g., clothing, temperature) may generate a different capacitance~\cite{grosse2017findingcommon}.

\subsection{Implementation}
We implemented the data processing stage with Python 3 and libraries for matrix computations such as NumPy/SciPy~\cite{virtanen2020scipy}. We used NetworkX~\cite{networkx} for the DFS-based resistor link selection. 
For the resistance optimization, we used Lcapy~\cite{hayes2022lcapy} and SymPy~\cite{meurer2017sympy} for the circuit simulation and symbolic computation and Pathos for multiprocessing.
The network layout adjustment is implemented with PyTorch~\cite{paszke2019pytorch}.

For the digital fabrication stage, we used Rhinoceros 7~\cite{rhino} as the CAD software and its programming environment, Grasshopper~\cite{grasshopper}, to computationally build 3D models.
We used a Prusa i3 MK3S+ 3D printer coupled with a Mosaic Palette Pro 2 to enable multi-material printing.
Our conductive filament is Protopasta's conductive PLA (\SI{1.75}{\milli\meter})~\cite{protopasta}. This filament is commonly available and provides a good balance of conductivity and resistivity to design a sensing network. The non-conductive filament can be any standard PLA filament. We used iSANMATE Wood Filament PLA+ (\SI{1.75}{\milli\meter})~\cite{woodpla} to emphasize color contrast between the nodes and links. For our microcontroller, we tested both the Arduino UNO R3 (\si{16\mega\hertz} CPU) and UNO R4 WiFi (\si{48\mega\hertz} CPU)~\cite{arduinouno}. Both have a 5V power source and a 2.5V logic threshold.
To constantly measure the time delays, we utilized signals from the microcontroller's digital I/O pins~\cite{baesupplmaterials}.

\section{Evaluations}
We evaluate our computational pipeline in three ways: (i) a quantitative scalability evaluation, (ii) three usage scenarios driven by three general visualization tasks, and (iii) an interview with six domain experts.

\subsection{Scalability Evaluation} 
We conduct a two-part scalability evaluation. First, we establish the practical performance limitations of the data processing stage (i.e., resistor network and network layout design) for networks of various sizes and confirm the effectiveness of our resistance optimization in determining a user's touch selection (\autoref{sec4:computational-scalability}). Second, we analyze how large of a network we can fabricate with our approach (\autoref{sec4:fabrication-scalability}).

\subsubsection{Computational Scalability}
\label{sec4:computational-scalability}

\textbf{Time complexity analysis.}
The DFS for resistor link selection is $\mathcal{O}(N (N {+} L))$ where $N$ and $L$ are the numbers of nodes and links, respectively.
The main computation for resistance optimization is the matrix inversion, which is required to solve the equation for each touched node (line 7 in \autoref{alg:resistance-opt}). This has $\mathcal{O}(N S^3)$ where $S$ is the number of involved symbols. 
The number of symbols is linearly correlated to the number of resistor path branches, $B$ (e.g., $B{=}2$ for \autoref{fig:circuit_network_design}c). We can thereby rewrite the time complexity as $\mathcal{O}(N B^3)$. The network layout process's time complexity depends on existing algorithms.
Our default layout is the Fruchterman-Reingold layout, which has $\mathcal{O}(N^2 {+} L)$.
Lastly, the network layout adjustment's time complexity varies based on which loss functions are employed.
$J_\mathrm{\scriptscriptstyle int}^\mathrm{\scriptscriptstyle res}$ has $\mathcal{O}(N^2)$ as it involves comparing all resistor links. Similarly, $J_\mathrm{\scriptscriptstyle int}^\mathrm{\scriptscriptstyle non\_res}$ and $J_\mathrm{\scriptscriptstyle int}^\mathrm{\scriptscriptstyle node}$ have $\mathcal{O}(L^2)$ and $\mathcal{O}(N^2)$, respectively. The time complexities for $J_\mathrm{\scriptscriptstyle pos}$ and $J_\mathrm{\scriptscriptstyle len}$ are almost negligible compared to the other loss functions. 
For each iteration performed by a NN, the network layout adjustment has $\mathcal{O}(N^2)$ at minimum as we must include $J_\mathrm{\scriptscriptstyle int}^\mathrm{\scriptscriptstyle res}$ and $J_\mathrm{\scriptscriptstyle int}^\mathrm{\scriptscriptstyle node}$. 
When using all the loss functions, the time complexity becomes $\mathcal{O}(N^2 {+} L^2)$.

\begin{figure}[tb]
	\centering
	\includegraphics[width=\linewidth]{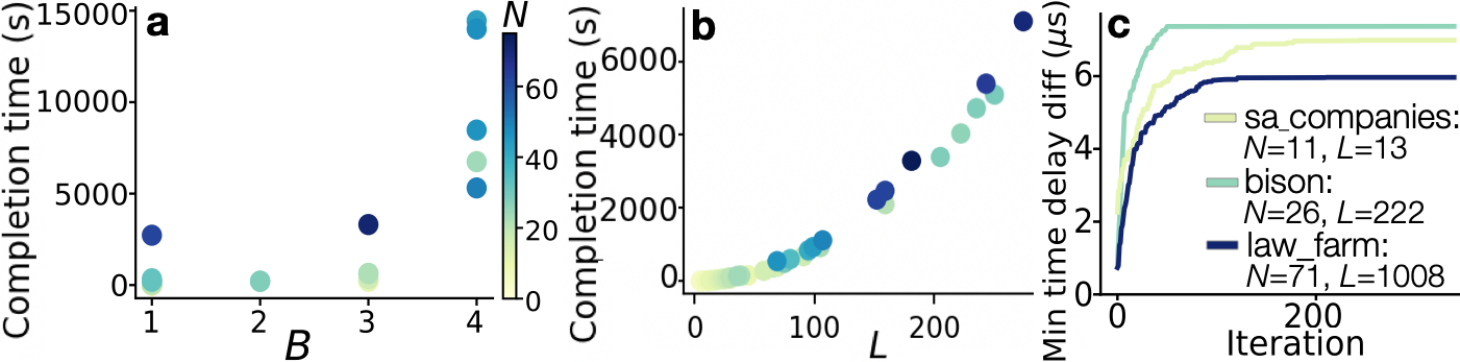}
    \caption{Computational evaluation results: (a, b) performance of resistance optimization and network layout adjustment; (c) progress of resistance optimization over iterations. Because of long completion time, we only show the results for networks with $B \leq 4$ in (a) and networks with $L \leq 300$ in (b). For (c), we show the results for three different networks.}
    \label{fig:comp-eval}


    \vspace{-1em}
\end{figure}

From this analysis, the network layout adjustment to avoid link intersection has the largest computational cost ($\mathcal{O}(N^2 {+} L^2)$).
The matrix inversion, which uses symbolic computing, also usually requires time-consuming computations even when the number of symbols is very small (e.g., \SI{1000}{\second} when $S{=}10$~\cite{hayes2022lcapy}).
Resistance optimization ($\mathcal{O}(N B^3)$) can be a bottleneck when a network dataset has a large number of path branches, $B$ (e.g., a network with many star shape sub-networks).

\textbf{Experimental evaluation.}
We used a MacBook Pro with 2.3 GHz 8-Core Intel Core i9 and 64 GB 2,667 MHz DDR4 (no GPU use). 
We collected networks from the graph-tool's dataset collection~\cite{peixoto_graph-tool_2014} and the Netzschleuder network repository~\cite{netzschleuder} to evaluate our pipeline with various real-world and synthetic networks.
Given current limitations on 3D printing resolution, we selected all networks with less than 100 nodes, resulting in 57 networks with the ranges of $N$: [4, 96] and  $L$: [5, 2539].
For layout adjustment, we measured the time needed to process a network with all loss functions.

As expected from the time complexity analysis, the resistance optimization and the network layout adjustment had longer completion times than other pipeline phases. The friendship network of New Guinea Highlands tribes ($N{=}16$, $L{=}58$) took \SI{0.7}{\milli\second} for resistor link selection, \SI{50}{\second} for resistance optimization, \SI{4}{\milli\second} for network layout, and \SI{300}{\second} for network adjustment.
From this bottleneck, we focus on evaluating the completion times of the resistance optimization and network adjustment.

The results of resistance optimization and network adjustment are shown in \autoref{fig:comp-eval}a and b. 
The $x$-axis reflects the most influential variable for their time complexity (e.g., $B$ for resistance optimization due to $\mathcal{O}(N B^3)$). The sequential colormap reflects $N$ as the secondary influential variable.
These results follow the theoretical time complexity. 
The resistance optimization finished in approximately 4 hours for a network with $N{=}47$ and $L{=}504$ ($B{=}4$).
The retwork layout adjustment completed in 2 hours for a network with $N{=}70$ and $L{=}274$.
While these completion times are nontrivial, it would take much longer and be significantly more error-prone to design appropriate resistance and layout by hand.

Lastly, we evaluated the quality of resistance optimization to examine its effectiveness in practically distinguishing touched nodes with a microcontroller. 
We initialized the resistance of resistor links randomly within a range of [\SI{50}{\kilo\ohm}, \SI{300}{\kilo\ohm}] and performed optimization.
\autoref{fig:comp-eval}c shows the transition of the minimum difference among the time delays. 
Optimization significantly increases the minimum difference (e.g., from \SI{0.8}{\micro\second} to \SI{7.4}{\micro\second} for the bison network) and quickly converges. As mentioned in \autoref{sec3:resistance_opt}, our resistance optimization is designed such that it maximizes the minimum difference of the time delays. 
An order of magnitude increased time delays significantly improves the sensitivity of node recognition. This improvement enables the Arduino UNO R4 to further distinguish touched nodes by the difference of 355 clock cycles (i.e.,  \SI{7.4}{\micro\second}) instead of 38 clock cycles (i.e., \SI{0.8}{\micro\second}).

\begin{figure}[tb]
	\centering
	\includegraphics[width=\linewidth]{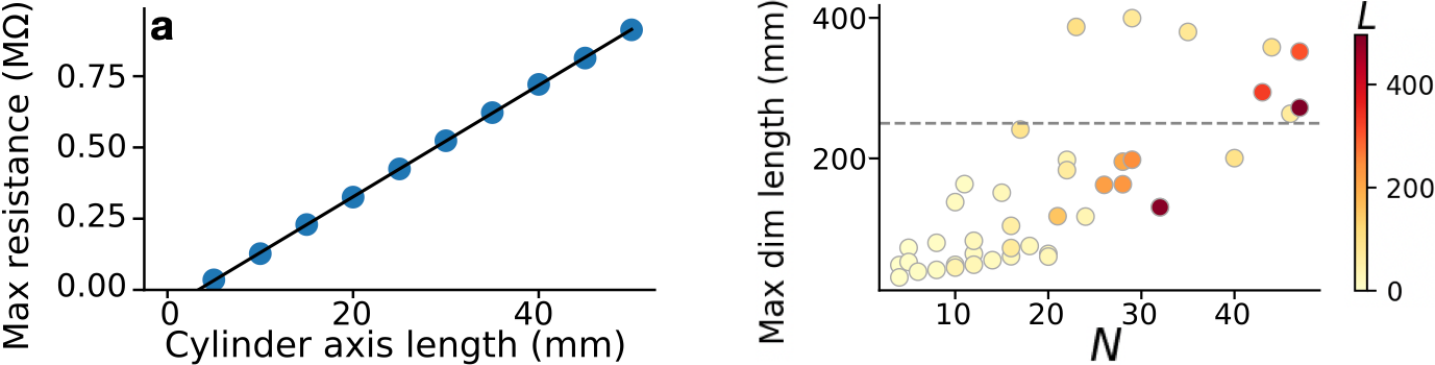}
    \caption{Fabrication scalability analysis: (a) the maximum resistance we can create for different cylinder lengths; (b) the longest dimension lengths required for printing space, where 250\,\SI{}{\textsf{\milli\meter}} corresponds to the maximum printable size using a Prusa i3 MK3S+.}
    \label{fig:fab-eval}

    \vspace{-0.5em}
\end{figure}

\subsubsection{Fabrication Scalability}
\label{sec4:fabrication-scalability}
To determine the largest network we can fabricate, we assume a 3D printer with a \SI{0.4}{\milli\meter} nozzle (standard for consumer FDM 3D printers) using Protopasta’s conductive PLA (\SI{1.75}{\milli\meter}).
We first measured the maximum resistance we can produce with the conductive traces using the serpentine trace pattern.
Because the maximum resistance depends on the cylindrical volume of a resistor link, we applied our resistor network embedding method to cylinders with different axis lengths and a fixed base radius of \SI{3}{\milli\meter}.
In \autoref{fig:fab-eval}a, we see a linear relationship between the maximum achievable resistance and the cylinder length.
 
Next, we identified the required resistance for sensing networks' resistor links to be able to distinguish each selected node.
To keep the evaluation concise, we assume the following: the sensing networks are paired with the Arduino UNO R4 (one clock cycle per \SI{21}{\nano\second}); 
the time delay to distinguish a touched node is at least a difference of \SI{2.1}{\micro\second} (100 clock cycles);
the involved capacitance for each touch selection is \SI{100}{\pico\farad} (a representative value for a human body~\cite{esd2010fundamentals}); and all resistor links are connected in series.
With these assumptions and \autoref{eq:voltage-transient}, the derived required resistance for each resistor link is \SI{30}{\kilo\ohm}.

We use this resistance value to derive the required print sizes to fabricate sensing networks.
As shown in \autoref{fig:fab-eval}a, the minimum cylindrical length to satisfy \SI{30}{\kilo\ohm} is \SI{17}{\milli\meter}.
We set the radius of each node to be \SI{6}{\milli\meter} (twice the base radius of a cylinder).
Following our network layout and adjustment processes, we laid out the networks such that their minimum resistor link length matched the identified cylinder length.
Note that we only used $J_\mathrm{\scriptscriptstyle int}^\mathrm{\scriptscriptstyle res}$ and $J_\mathrm{\scriptscriptstyle int}^\mathrm{\scriptscriptstyle 
 node}$ to avoid intersections of conductive materials.
We then obtained the maximum distance of all possible pairs of nodes to compute the required length for the printing area.

Based on these computations (\autoref{fig:fab-eval}b), we estimate that the largest printable sensing network size is between 20--30 nodes when using a consumer 3D printer such as the Prusa i3 MK3S+. This result implies that the scale of our approach is currently limited by the size of the 3D printer rather than the computational complexity of the algorithms.

\subsection{Usage Scenarios}
\label{sec4:usage-scenarios}
We present three example usage scenarios to illustrate how our sensing networks can be used in practice.
These usage scenarios motivate how selection can support three general visualization tasks: exploration, explanation, and analytic provenance.
The scenarios are contextualized in the educational context of a student exploring disease spread using a social network, and are driven by the concept of \textit{exploranation}~\cite{ynnerman2018exploranation} (i.e., how exploratory and explanatory visualization techniques can inform each other). The exploratory case demonstrates how our approach aligns with emerging uses of tangible 3D input devices~\cite{besanccon2021state,huang2017gesture}. 

Various studies confirm the benefits of using 3D input devices, such as reducing mental effort, compared to the traditional keyboard and mouse setup~\cite{besanccon2017mouse, satriadi2022tangibleglobes, huang2016embodied}. The explanatory case aligns with how physical models have also been used to richly explain abstract concepts (e.g., visualization~\cite{huron2014constructive, bae2022cultivating}, biology~\cite{ang2019physicalizing}) for educational purposes and general science communication~\cite{hansrosling2014,Schönborn2016nano}. 
The third usage scenario emphasizes our network physicalization's capabilities as a sensor. Building upon literature on analytic provenance~\cite{fujiwara2018concise,ragan2016chacterizingprovenance}, our networks can richly and passively capture how users are interacting with the data physically, serving as a tangible data log for their physical navigation of the data.

We derived these scenarios from discussions with subject matter experts who use networks in their everyday practice (e.g., chemistry, computational biology, network science). The usage scenarios summarize core functionality that experts currently rely on using alternative solutions. They also demonstrate the broad utility of our sensing network by enabling serial communication with other devices. Each scenario is based on the parameters described during our interviews (\autoref{fig:disease}), and uses the network physicalization in \autoref{fig:sec3:network-overview}. Details of these discussions are stated in \autoref{sec4:expert-feedback}. See the supplemental video for demo~\cite{baesupplmaterials}.

\subsubsection{Exploration and Discovery}
\label{sec4:exploratory}

\textbf{Background.} 
Students can learn about viral transmission through social networks, which model the connections between students.
A professor prepares a synthetic social network ($N=20$, $L=40$) where each student node has a set of attributes such as age, gender, and number of social connections. The network is represented in 2D (digital) and 3D (physical). Students apply graph theory concepts to analyze the social interactions that determine disease dynamics. 

A student named Alice assesses the network to identify individuals who have a high potential to spread the disease. Her task is to determine which centrality measure would lead to the best vaccination strategy. She looks at the 2D and 3D representations of the same network. With both representations, she immediately notices there are two distinct clusters, but the link intersections of the righthand cluster are difficult to distinguish on screen. 
Alice connects the physical model to a desktop visualization for navigation and initial exploration. She wants to find Node 2 who is identified as the one infected person. The desktop currently displays all nodes and their ID. She notices Node 2 is within the right cluster. She \textbf{filters} out all nodes in the left cluster by selecting each node. These filtered nodes turn grey and their opacity reduces.

\begin{figure}[tb]
	\centering
	\includegraphics[width=\columnwidth]{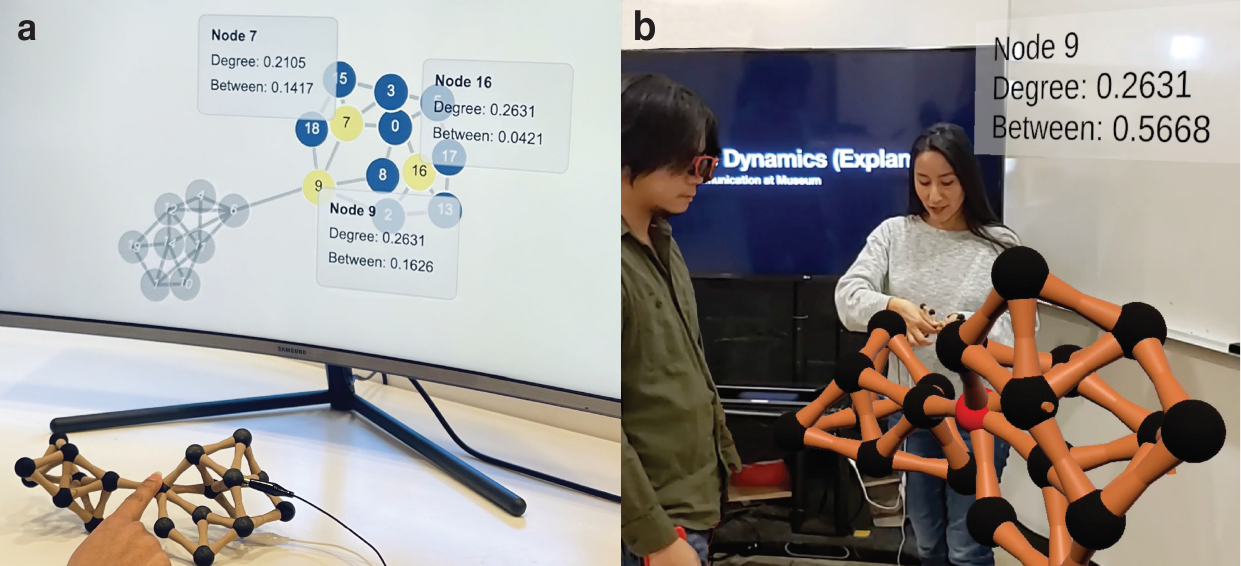}
    \caption{Three usage scenarios (\autoref{sec4:usage-scenarios}) with a sensing network physicalization ($N=20$, $L=40$; see \autoref{fig:sec3:network-overview}). (a) Exploratory: Alice analyzes the network using graph centrality and concludes Node 9 should be vaccinated; (b) Explanatory: Alice touches Node 9 to explain her findings. Network is serially connected to audiences' AR glasses and mobile AR applications. Note: displayed degree and betweenness centralities are normalized by dividing by the highest possible values for this network.}
    \vspace{-1em}
    \label{fig:disease}
\end{figure}

Alice double-taps all the nodes within the righthand cluster to \textbf{inspect local details} for their centrality measures. A tooltip appears. She analyzes the degree centrality of each node. Nodes 7, 9, 16 have high degree centrality, representing individuals with many connections in the networks. Alice holds onto these respective nodes for three seconds to \textbf{highlight} them in yellow  (\autoref{fig:disease}a).
She then analyzes the betweenness centrality of the highlighted nodes, which represents the extent to which individuals lie on the shortest paths between pairs of other individuals in the network. She \textbf{compares} the betweenness centrality among the three nodes. From the tooltip, Node 9 has the highest betweenness centrality. She concludes that Node 9 should be vaccinated.

\subsubsection{Explanation and Communication}
\label{sec4:explanatory}

Alice presents her findings (i) to her classmates and teachers remotely and (ii) during a live presentation at a science center. 

\textbf{Remote presentation.}  
Remote presentations lack shared physical context, making deictic gestures, such as pointing to items of interest, difficult to interpret~\cite{verhulsdonck2007issues}. To provide context for her findings, Alice uses her model as a physical prop during a video call with her class to present her findings. She can use deictic gestures to indicate key nodes and links during her presentation by pointing, with the physical model providing a shared context between Alice and her remote audience. Alice can also use the desktop visualization as a digital twin of the network~\cite{hu2022remote} and use the sensing network to interact with the twin. As she touches different nodes, the sensing network allows the twin to highlight those nodes in red and display additional information about the corresponding student in the network. When she releases the node, the information disappears, allowing her to continue her discussion.

\textbf{Science communication in public museums.} 
Science communication at museums or local science centers bridges researchers and the general public. 
Visitors can engage in research data with algorithms and software similar to what was used by researchers. 
However, public displays impose installation challenges in terms of reliability and interaction. To amend this challenge, museums often use touch surfaces that enable robust and intuitive multi-user interaction~\cite{ynnerman2016interactive}. Despite the documented success of touch interfaces, tangible interfaces and physicalizations can better scaffold knowledge and promote learning~\cite{Schönborn2016nano}.
Alice uses her physicalization in a workshop presentation at the local science center. 
AR headsets and mobile AR are used to display augmented information on the network model. She presents her findings and lets the participants handle the model and study the responses to the augmented information. 
Rather than relying on the cameras on headsets and phones for interaction, which suffer from occlusion, she uses her digital twin configuration to bridge the physical model with the audience members' devices. As Alice is walking through her data exploration (\autoref{fig:disease}b), the sensing capabilities of the network inform the headsets of the different node selections and display the corresponding node information. Audience members can see her node selections even when the model is outside of the headset's and phone's field of view.

\subsubsection{Provenance and Documentation}
\label{sec4:provenance}
As Alice explores and explains her data, she physically interacts with the network. What nodes she touched, for how long, and in what order can all be logged to replay her physical interactions with the data~\cite{fujiwara2018concise}. The innate sensing capability challenges conventional approaches to capture interactive data exploration (i.e., data sensemaking). Based on Alice's exploration (\autoref{sec4:exploratory}), her professor analyzes her logged interactions to understand her data sensemaking process. He wants to use this information to better inform his teaching curriculum. He visualizes the logged data as a timeline visualization and a heatmap. 

In the visualizations, he notes that Alice kept ``bookmarking''~\cite{jansen2013evaluating, drogemuller2021haptic2Dvis} Node 9 by leaving her finger in place. The lingering action confirms that Alice was thoroughly exploring the righthand cluster. He notices that she only engaged with the left cluster only in the beginning of her exploration. He plans to note that if Alice did not immediately filter out the left cluster's nodes, she would have noticed that Node 6 also has a high betweenness centrality and thus should also be vaccinated. 

\subsection{Expert Discussion on Utilization}
\label{sec4:expert-feedback}
We discussed our sensing networks with six domain experts (1 female, 5 males; age reported in bins of 26--55 years). Their expertise varied in AR/VR (3), material science (1), computational biology (1), and high-dimensional data (1).Four are working professionals in academia or national labs. Two are Ph.D. students. Two interviews were conducted remotely and four were in person. See the supplementary materials~\cite{baesupplmaterials} for more details.

We first explained the basic principles of the pipeline (\autoref{sec3:basic_principles}) and demonstrated the networks either in-person or remotely. Within the physical condition, we asked the domain experts to interact with the physical network themselves and demonstrated how it can be connected to other devices. Within the remote condition, we demonstrated the sensing capability by screen-sharing the model and playing videos we captured in advance.  We then asked questions designed to elicit both insights into potential uses as well as preliminary perceptions of the pipeline and resulting networks.

Overall, the domain experts reacted positively to our produced networks, with E6 (an AR/VR developer) saying, ``\textit{Currently, there are only a few ways objects can meaningfully respond to touch, and your technique is changing that [paradigm].}'' Five of the six experts noted the sensing network's potential for supplementing existing exploratory workflows. E1 (expert in uncertainty in high-dimensional data) expressed how it could potentially be used as a parameter tuning control for high-dimensional spaces (e.g., producing an interactive Morse-Smale Complex~\cite{gerber2010morse}). 
E1 felt she could ``\textit{play with the graph of the [high-dimensional] parameter space to [move] to new locations in the output space, [which can] give us a better understanding of how parameters are related and ways to navigate the parameter space.}'' E2, a material scientist, uses CAVEs to immersively see static 3D molecular compounds (e.g., solar cells). However, he often faces difficulties interacting with the CAVE to filter certain sub-components to visualize further. He envisioned that the networks can possibly act as  interactive ``motifs'' (similar to graphlets~\cite{shervashidze2009efficient}) within the large molecular polymers he analyzes. The network's sensing capabilities would enable more intuitive interactions, while the pipeline would enable him as a domain scientist to readily produce polymer physicalizations on demand.

Experts who develop AR/VR provided similar comments on how this can help supplement existing interaction workflows for immersive spaces. E4-E6, all AR/VR developers, stated that the ability to produce haptic objects that share the form of critical virtual objects and can act as input sensors will benefit the field. E6 specified that the biggest issues with current haptic proxies stem from how they are either completely passive or require too much instrumentation to make them interactive. Our technique would enable developers to more easily produce responsive controllers shaped like networks or other hub-and-strut forms, which can complement immersive analytics tools.

E3 (a computational biologist) and E6 provided useful feedback for future work. Despite explaining the scalability of the technique, E3 noted that,``\textit{I'm not sure if I want to even hold a 50+ node network}''. This comment aligns with findings that the physical scale of a  physicalization needs to be chosen carefully for ease of manipulation and representation legibility~\cite{lopez2021scaling}. E6 pointed out how the current network could be more powerful by being modular and reconfigurable.

\section{Discussion}
\label{sec5:discussion}
\textbf{Need for computational fabrication processes}.
By concurrently generating data representations (form) alongside sensing capabilities (interaction), our pipeline addresses several low-level engineering challenges associated with physicalization production (e.g., layout, circuit design, and sensor integration) for networks and any visualization using a hub-and-strut structure (see \autoref{sec3:network-layout-overview}).
This design paradigm shift enables integrating sensing capabilities without disturbing a physicalization's form and can help achieve more complex physicalizations. 

Past work in visualization authoring tools (e.g., D3~\cite{bostockd3} and Vega-lite~\cite{satyanarayan2016vega}) parallels our goal to make it easier to produce interactive data representations. 
The power of these toolkits is their ability to concurrently consider interactions and visual representations, similar to our pipeline. However, prior to these tools, developers used systems designed for general graphics applications, which lacked support for designing interactive visualizations. Consequently, developers had to develop custom solutions for common operations (e.g., mapping data to visual elements and event handlers).  
Digital toolkits reduced the time and labor associated with visualization development, leading to readily reusable and reliable components. Toolkits for physicalizations have the potential to achieve similar goals, enabling developers to focus more on \textit{what} analytical scenarios they want to explore with a physicalization rather than \textit{how} to build one. We contend that our pipeline can provide an important step towards more robust toolkits for interactive physicalization. As a result, this work lowers the barrier of entry for people to incorporate physicalizations into their work.

\textbf{Enabling output.} Interactive objects receive input (e.g., from touch) and produce output (e.g., light, sound, color change) in a controlled manner. Our sensing network currently addresses the first part of the interaction loop by responding to touch inputs and can serve as a foundational step to truly realize interactive physicalizations.
One way to enable display output in our sensing networks is to integrate trace routing approaches for 3D printed optical fibers~\cite{willis2012printedOptics} or color-changing dyes~\cite{jin2019photo-chromeloen}. Color is one of the main visual channels when representing data, and the existing work in this space (e.g., visualizations with color-changing inks~\cite{Patnaik2022}) suggests its potential for physicalizations. Another possibility is to explore sound output with 3D printed speakers~\cite{ishiguro20143d}, which lends well to existing research efforts on sonification~\cite{hermann1999listen}. 

\textbf{Working alongside other visualizations.} 
Our computational pipeline currently supports selection by touch through capacitive sensing and computational inference. Though limited, select is a fundamental interaction primitive (\autoref{sec2:select}) that enables designers to implement more complex interactions at the application layer (see \autoref{sec4:usage-scenarios}). Currently, the output of our interaction loop is achieved through digital visualizations. Though future work should investigate how to enable physical output, experts saw value in this current digital-physical paradigm. This value stems from how our experts work with different media (e.g., CAVEs, desktops, AR/VR headsets) and how our sensing network physicalizations can be easily integrated with these different devices through serial communication. This integration allows systems to simultaneously leverage the responsiveness of digital displays with the intuitive interactions and cognitive benefits of physicalizations.

Nonetheless, future work is necessary to see which display methods (e.g., tabletop, shape-changing, AR/VR) best complement interactive physicalizations. In turn, this hybrid modality can shed a more critical perspective on analyzing the strengths and weaknesses of digital and physical approaches, possibly leading to a stronger post-WIMP (windows icon menu pointer) paradigm~\cite{roberts2014beyond, jacob2008rbi}. 
Future work should continue to investigate how to integrate input and output modalities into physicalizations by concurrently considering form and interactivity.

\section{Future Work}
\label{sec6:future-work}

\textbf{Improving scalability.} 
Our fabrication process is the primary bottleneck for scalability (\autoref{sec4:fabrication-scalability}). While FDM 3D printers are more affordable and approachable to end-users, they have a smaller build volume, slower speed, and less printing resolution compared to other processes~\cite{ngo2018additive}. 
For example, the network shown in \autoref{fig:sec3:network-overview} ($N{=}20$, $L{=}40$) took 50 hours to print. A multi-nozzle 3D printer~\cite{ vasquez2020julibee} could print much larger networks at a faster rate while having more precision in the design of conductive traces. This process can help fully explore the benefits of larger, physical networks (e.g., extracting depth information) while still supporting interactivity. To support higher precision and larger scale, our algorithms' performance must also improve. We could improve resistance optimization by simplifying resistor links connected in parallel. This would require future research into distributing gradients for the combined resistor to the original resistor links. 

\textbf{Improving calibration.} Currently, our sensing technique requires manual calibration for each printed network and user due to the differences in capacitance. In addition, external environmental factors (e.g., temperature, surrounding materials) can affect the capacitive sensors. One way to partially automate this process is to measure a user's capacitance as they touch a few nodes and compute the corresponding time delays with the circuit simulator used for resistance optimization. However, this optimization would require high precision while printing conductive traces to ensure the sensing network and the circuit simulator have the same resistance for each resistor link.

\textbf{Improving design flexibility.}
Our approach currently does not support link selection. To support concurrent node and link selection, we need to imbue both surfaces with high conductivity while creating conductive traces within.
These constraints bring additional complexity that will need to be addressed by the resistance optimization, layout adjustments, and fabrication process. We also plan to support different shapes for nodes and links to allow more flexible physical encodings, such as arbitrarily-shaped nodes, curved and bundled links. These designs can take further advantage of the computational process of 3D printing, which can help physicalizations represent high-dimensional, multivariate datasets.
While the nodes of our sensing networks are uniform spheres, we can add additional visual channels by varying their size, color, texture, and other physical and visual properties. 
\section{Conclusion}
We introduce a computational design pipeline to 3D print network physicalizations with integrated sensing capabilities. This pipeline inputs network data, computes the internal circuitry for capacitive sensing, adjusts the layout to support fabrication, and produces a sensing network physicalization. Our methodology introduces a new design paradigm by concurrently considering form and interactions for physicalizations. By automating low-level hardware design challenges in favor of application-level implementations, our approach can create more complex and powerful tools for designing physicalizations. With this new design paradigm, we can produce generalizable techniques that lower the barrier to physicalization research, creation, and adoption.
\balance
\acknowledgments{
This research is sponsored in part by the U.S. National Science Foundation through grants IIS-2040489, IIS-1764089, IIS-2320920, IIS-1933915, IIS-2233316, the Knut and Alice Wallenberg Foundation through grant KAW 2019.0024, and the CU Boulder Engineering Education and AI-Augmented Learning Interdisciplinary Research Theme Seed Grant.
}

\bibliographystyle{abbrv-doi-hyperref}
\bibliography{template}

\end{document}